\documentclass[conference]{IEEEtran}
\IEEEoverridecommandlockouts
\usepackage[T1]{fontenc}
\usepackage{graphicx}
\usepackage[nobreak]{cite}
\usepackage{url}
\usepackage{xspace}
\usepackage{xcolor}

\usepackage[fleqn]{amsmath}
\usepackage{newtxmath}

\makeatletter
\let\MYcaption\@makecaption
\makeatother
\usepackage[font=footnotesize]{subcaption}
\makeatletter
\let\@makecaption\MYcaption
\makeatother

\usepackage{cleveref}
\crefname{figure}{Fig.}{Figs.}
\Crefname{figure}{Figure}{Figures}
\crefname{table}{Table}{Tables}
\crefname{section}{Section}{Sections}

\usepackage{framed}
\newcommand{\Conclusion}[1]{\begin{framed}\noindent #1\end{framed}}

\newcommand{\Heading}[1]{\textbf{#1.}}
\newcommand{\RQ}[1]{\textit{RQ}${}_{\mathrm{#1}}$}

\newcommand{\Tuple}[1]{\langle #1\rangle}
\def\Eq{\mathit{eq}}
\newcommand{\Pt}[2]{v^{#1}_{#2}}
\newcommand{\Edge}[4]{\Tuple{\Pt{#1}{#2},\Pt{#3}{#4}}}

\newcommand{\Fix}[2]{({#1},{#2})}
\newcommand{\Action}{\mathit{action}}

\def\Dfix{\mathit{diff}\!_\mathit{fix}}

\begin{document}

\title{Toward Interactive Optimization of Source Code Differences: An Empirical Study of Its Performance}

\author{
  \IEEEauthorblockN{Tsukasa Yagi}
  \IEEEauthorblockA{
    \textit{School of Computing}\\
    \textit{Tokyo Institute of Technology}\\
    Tokyo 152--8550, Japan\\
    tsukasa@se.c.titech.ac.jp}
  \and
  \IEEEauthorblockN{Shinpei Hayashi}
  \IEEEauthorblockA{
    \textit{School of Computing}\\
    \textit{Tokyo Institute of Technology}\\
    Tokyo 152--8550, Japan\\
    hayashi@c.titech.ac.jp}
}

\maketitle

\pagestyle{plain}
\thispagestyle{plain}

\begin{abstract}
A source code difference (diff) indicates changes made by comparing new and old source codes, and it can be utilized in code reviews to help developers understand the changes made to the code.
Although many diff generation methods have been proposed, existing automatic methods may generate nonoptimal diffs, hindering reviewers from understanding the changes.
In this paper, we propose an interactive approach to optimize diffs.
Users can provide feedback for the points of a diff that should not be matched but are or parts that should be matched but are not.
The edit graph is updated based on this feedback, enabling users to obtain a more optimal diff.
We simulated our proposed method by applying a search algorithm to empirically assess the number of feedback instances required and the amount of diff optimization resulting from the feedback to investigate the potential of this approach.
The results of 23 GitHub projects confirm that 92\% of nonoptimal diffs can be addressed with less than four feedback actions in the ideal case.
\end{abstract}

  
\section{Introduction}\label{s:introduction}
Code review is a process in which developers manually inspect the source code to identify defects and enhance the quality of the software \cite{McIntosh2015}.
This process requires significant effort, and therefore, a modern code review, which is a typical approach for code review that inspects modified parts of the source code \cite{Bacchelli2013}, is employed.

Understanding changes is crucial to properly conduct code reviews \cite{Yida2012}.
Review tools display changes made to the source code using line differences\cite{Bacchelli2013}.
These line differences facilitate the developers' understanding of the changes\cite{Yida2012}.

However, existing methods for generating line differences may not accurately identify changes in the source code \cite{Nugroho2020}.
Moreover, displaying differences that represent logically coherent changes is preferable for a better understanding of changes \cite{Ram2018, Barnett2015}.
However, there may be instances where the grouping of change is not optimized.
Further, nonoptimal differences can impact tools that provide additional information based on line differences \cite{Canfora2009, Nugroho2020}.

We propose an interactive optimization method to address the issue of unoptimized line differences.
In this method, users review the output line differences and provide feedback where change identification is deemed inappropriate.
Subsequently, users can obtain more optimized differences by generating new line differences that incorporate this feedback.

A significant divergence from other automated difference generation methods is that this method requires human manipulation.
Therefore, evaluating whether the effort required for feedback operations is realistic is necessary to discuss the practicality of this method.
However, the direct quantification of effort with human subjects requires significant cost, which is beyond the scope of this paper.
Instead, we preliminary attempt to quantify the optimization performance from two aspects: the amount of feedback required to obtain optimal differences and the amount of improvement in differences via feedback.
We formulated the process of repetitive feedback in the interactive optimization as a graph search problem and simulated it by applying a search algorithm.

The source code changes from 23 projects hosted on GitHub were used as subjects for the study.
The feedback performance was evaluated based on the simulation results.
The obtained results indicate that feedback was provided 1.73 times (on average) for optimization in the ideal case; 4.87 locations were optimized simultaneously with an instance of feedback.
The amount of optimization by average feedback was 68\% compared to that of the ideal case.
These results suggest that, in the ideal and average cases, this method can optimize differences without much effort.
Therefore, it is meaningful to continue research towards practical application.

The main contributions of this paper are as follows:

\begin{itemize}
  \item Proposal of an interactive optimization method for differences.
  \item Empirical study of the potential performance of the interactive optimization method, which reveals the optimization performances in ideal and average cases.
\end{itemize}

The remainder of this paper is outlined as follows:
\Cref{s:motivation} discusses the motivation for this study.
\Cref{s:technique} explains the method for generating feedback-based differences.
\Cref{s:sim} describes the simulation method used for the empirical study.
\Cref{s:empirical} discusses the results of the optimization performance investigation.
\Cref{s:relatedwork} summarizes related work on code differencing techniques.
Finally, \Cref{s:conclusion} outlines the contents of this paper and future challenges.

\begin{figure*}[tb]\centering
  \begin{subfigure}{\linewidth}\centering
    \includegraphics[width=0.8\linewidth]{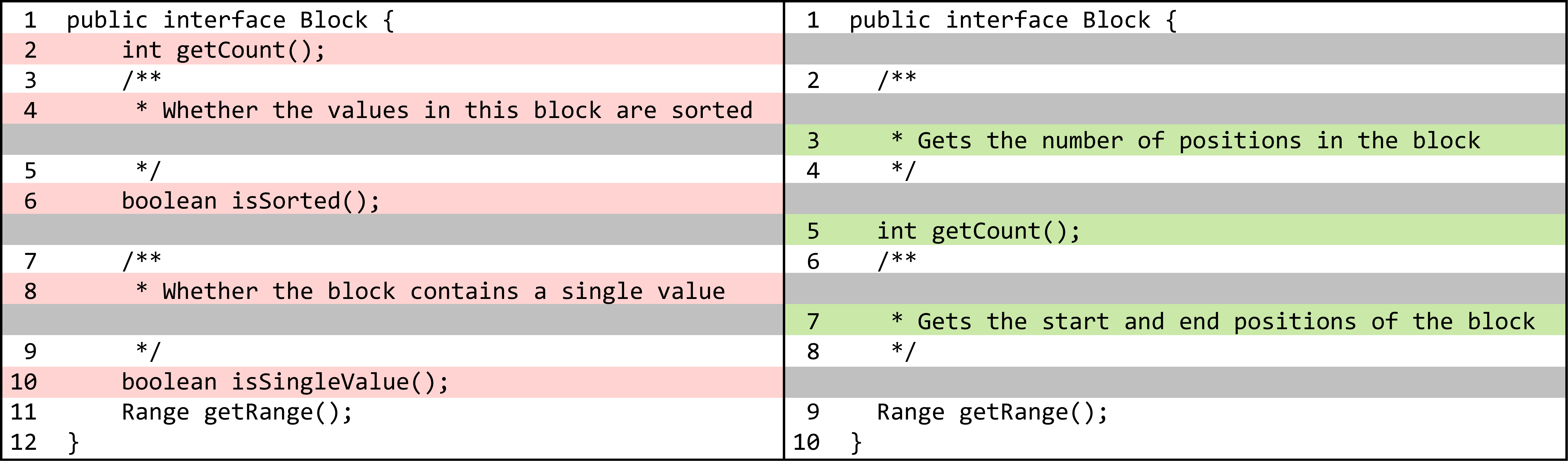}
    \caption{Nonoptimal diff.}\label{f:badDiff}\vspace{0.7em}
  \end{subfigure}
  \begin{subfigure}{\linewidth}\centering
    \includegraphics[width=0.8\linewidth]{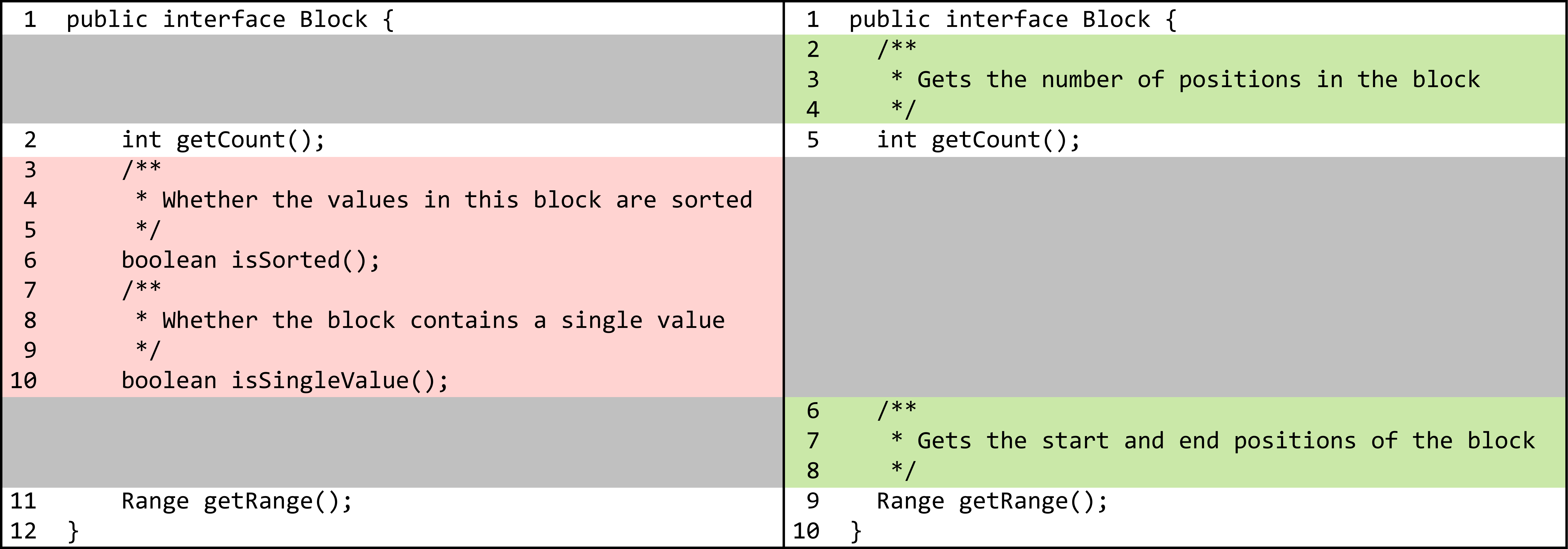}
    \caption{Optimal diff.}\label{f:goodDiff}
  \end{subfigure}
  \caption{Two diffs from the same source code pair.}\label{f:diff}
\end{figure*}

\section{Motivation}\label{s:motivation}   

Code review is the process in which an individual besides the author manually inspects the source code to identify defects and improve the maintainability of software.
The modern code review method reduces the workload by utilizing tools and focusing the inspection on changes made to the source code \cite{Bacchelli2013}.
A code review is conducted in numerous organizations, and several empirical studies have demonstrated its effectiveness in enhancing code quality \cite{McIntosh2015, Sadowski2018, Yida2012}.

The greatest challenge in code reviews is understanding the changes \cite{Bacchelli2013}.
Accurately comprehending changes leads to more effective code reviews \cite{Yida2012}.
Numerous code review support tools offer source code difference (diff), which developers use to understand modifications.
For example, GitHub computes and displays line diffs, ensuring that they are minimized as much as possible, similar to the output produced by Myers' algorithm \cite{Myers1986}.

However, diffs generated by existing methods may not always be optimal for the user.
Although diffs correctly describe the changes in calculations, there are instances where they do not accurately reflect the modifications made by developers \cite{Nugroho2020}.
Moreover, concise and logically cohesive diffs are preferred to effectively understand the changes \cite{Ram2018,Barnett2015}.
In nonoptimal diffs, cohesion may be compromised.

\Cref{f:diff} shows different source code diffs for the same changes: the removal of two methods and the addition of comments to the remaining methods.
\Cref{f:badDiff} depicts the diff generated by the Myers' algorithm\cite{Myers1986}, which is a widely used method, whereas \cref{f:goodDiff} is a diff with improved readability and understandability.

\cref{f:badDiff} is not optimized and is inconsistent with the intent of the change in several points.
For example, because the developer has not modified the \texttt{getCount} method, as illustrated in \cref{f:goodDiff}, the method must correspond between the two revisions; however, it does not in \cref{f:badDiff}.
Furthermore, comments are added to the \texttt{getCount} and \texttt{getRange} methods of the new version.
However, the diff in \cref{f:badDiff} correlates the comments between the new and old versions, failing to reflect the changes made by the developer accurately.
Existing methods for generating diff can produce such confusing diffs.
In fact, not only Myers but also other line differencing approaches, e.g., Histogram, cannot generate the diff shown in \cref{f:goodDiff}.
Currently, to our knowledge, no algorithm is known to consistently produce optimal diffs.

One may think that code diffs are just outcomes of automated differencing tools and not the target of manual optimization.
However, we believe that developers often encounter a scenario where the correction of diffs is useful.
For example, developers submit a source code diff to a mailing list or an issue tracking system of an open-source software project as a patch~\cite{patch-flow}.
The patches are the targets of the reviews from other developers.
Improving the understandability of a diff by a patch developer can reduce the time spent by multiple reviewers.
In this context, improving the understandability by correcting a diff must be a sufficiently low effort task.
However, it may be difficult or time-consuming to force developers to provide the correct mapping of lines when correcting a diff.
Therefore, a technique that can enable developers to provide a diff using an input easy to prepare without specifying the desired outcome is required.

\section{Interactive Optimization}\label{s:technique}

This paper proposes an interactive optimization for source code differences (diffs) as a solution to a problem that is difficult to solve with existing automatic diff generation.

\subsection{Preliminary: Edit Graph and Difference} \label{s:preliminary}

An \emph{edit graph} \cite{lcs-diff,Myers1986}\footnote{The definition in this paper does not strictly follow the ones in these references; however, it is an essentially equivalent one.} is an oriented graph $G = (V, E)$ consisting of sets of nodes $V$ and edges $E \subset V \times V$ where
\begin{align*}
    V =&\, \{\Pt{i}{j} \mid 0 \leq i \leq N \land 0\leq j \leq M\},\\
    E =&\, \{\Edge{i-1}{j}{i}{j} \mid 1 \leq i \leq N \land 0 \leq j \leq M \} \cup {}\\
       &\, \{\Edge{i}{j-1}{i}{j} \mid 0 \leq i \leq N \land 1 \leq j \leq M \} \cup {}\\
       &\, \{\Edge{i-1}{j-1}{i}{j}) \mid 1 \leq i \leq N \land 1 \leq j \leq M \land \Eq(i,j)\}.
\end{align*}
Here, $N$ and $M$ denote the numbers of lines of code in the old and new versions of the source code, respectively.
Node $\Pt{i}{j}$ expresses a state where the $i$-th and $j$-th lines of old and new versions of source code are read, respectively.
The edges are in between adjacent nodes.
We use three types of edges: \emph{horizontal}, \emph{vertical}, and \emph{diagonal}.
A horizontal edge $\Edge{i-1}{j}{i}{j}$ indicates that the $i$-th line of the old version of code was not associated with the lines in the new version and skipped (regarded as \emph{deleted}).
A vertical edge $\Edge{i}{j-1}{i}{j}$ indicates that the $j$-th line of a new version of the code was not associated with the lines in the old version and skipped (regarded as \emph{added}).
A diagonal edge $\Edge{i-1}{j-1}{i}{j}$ indicates that $i$-th line in the old version was associated with $j$-th line in the new version.
The predicate $\Eq(i,j)$ holds if and only if $i$-th line in the old version is the same as $j$-th line in the new version.
\begin{figure}[tb]\centering
  \includegraphics[width=\linewidth]{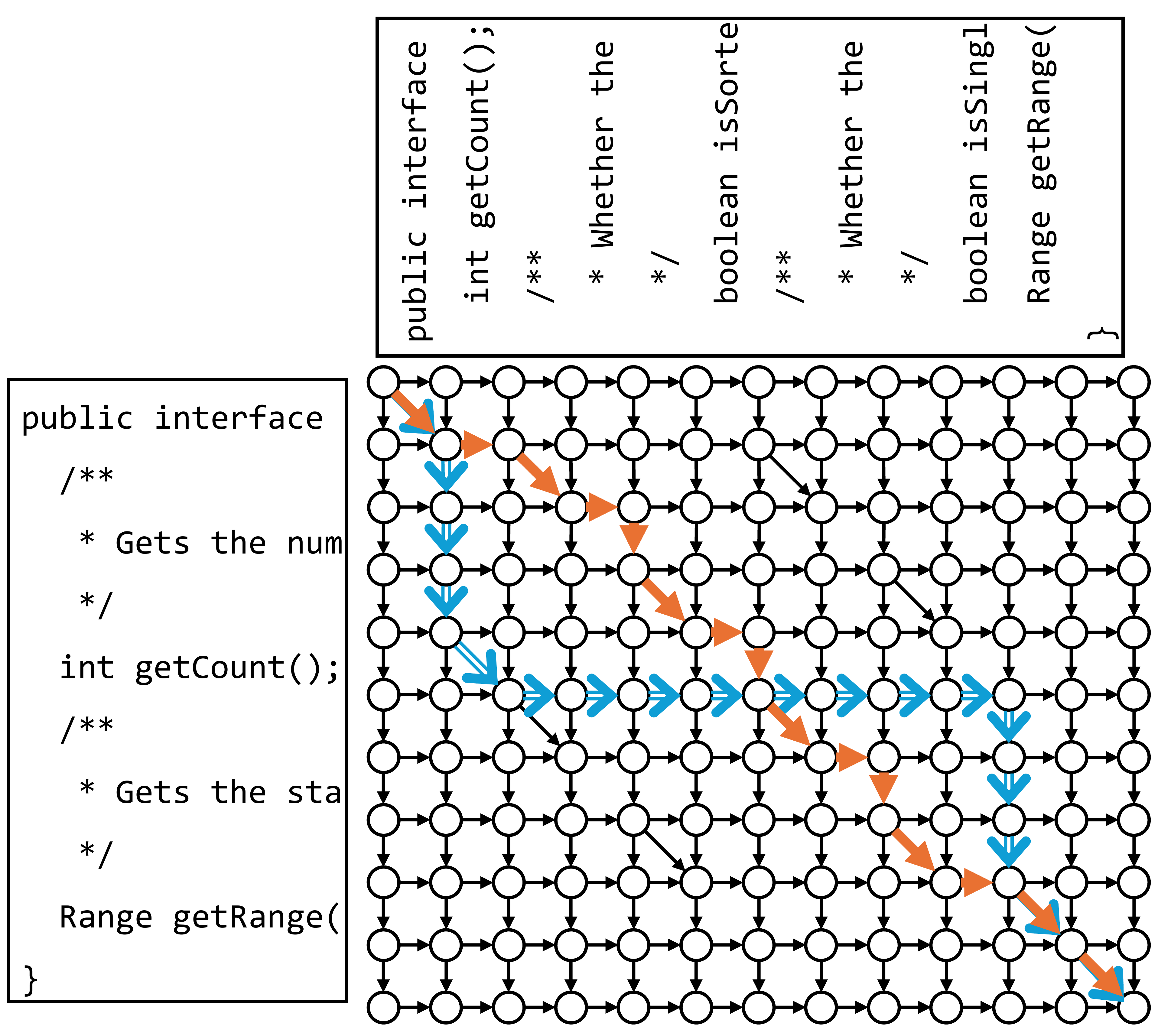}
  \caption{Example of an edit graph.}\label{f:editgraph}
\end{figure}
An example of an edit graph is shown in \cref{f:editgraph}, comparing the two versions of source code shown in \cref{f:diff}.

\begin{figure*}[tb]\centering
  \includegraphics[width=0.62\linewidth]{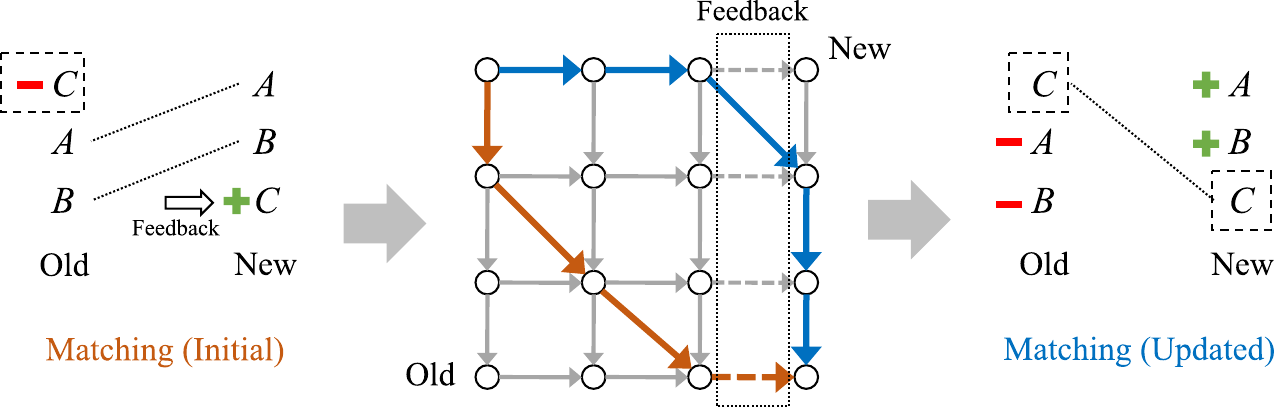}
  \caption{Overview of the interactive optimization.}\label{f:method-overview}
\end{figure*}
\begin{figure*}[tb]\centering
  \begin{subfigure}{0.23\linewidth}\centering
    \includegraphics[width=0.9\linewidth]{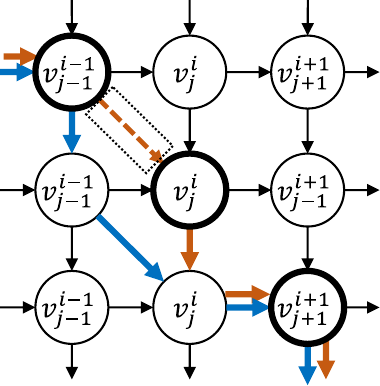}
    \caption{Mismatched line.}\label{f:mismatch}
  \end{subfigure}
  \begin{subfigure}{0.23\linewidth}\centering
    \includegraphics[width=0.9\linewidth]{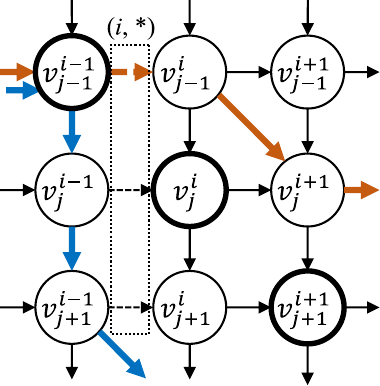}
    \caption{Old-orhphan line.}\label{f:old-orphan}
  \end{subfigure}
  \begin{subfigure}{0.23\linewidth}\centering
    \includegraphics[width=0.9\linewidth]{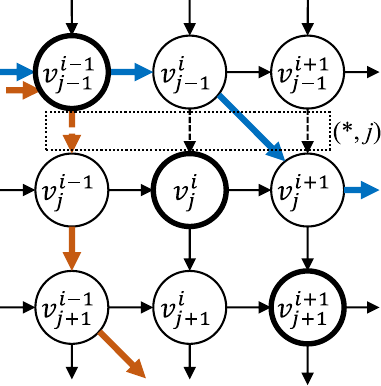}
    \caption{New-orphan line.}\label{f:new-orphan}
  \end{subfigure}
  \caption{Types of feedback actions.}\label{f:feedback}
\end{figure*}

A \emph{path} from node $\Pt{0}{0}$ to node $\Pt{N}{M}$ indicates a mapping between lines in the old and new versions, i.e., a \emph{diff}.
A diff $d \in D\,(\subseteq 2^E)$ can also be represented as a set of edges used to form the path.
We use one of the shortest paths of the graph as a preferable difference.
A set of orange-colored thick edges in \cref{f:editgraph} represents the shortest path in the graph, which corresponds to the difference shown in \cref{f:badDiff}.
The distance of this shortest path is 15.
The path includes five horizontal edges, which means that the difference includes the deletion of five lines.
Calculating the shortest paths on an edit graph is equivalent to calculating the \emph{longest common sequence}~(LCS) of the lines of code in the old and new versions.
If more than one shortest path is found, one of them is selected.

Many algorithms to calculate the LCS between lines have been proposed.
For example, a simple algorithm based on dynamic programming works under $O(NM)$.
More efficient algorithms are also proposed, e.g., $O(ND)$ \cite{Myers1986}.

\subsection{Basic idea}

The basic idea of the proposed technique is simple, namely, users modify the given edit graph based on their feedback.
The usage process of the proposed technique is shown in \cref{f:method-overview}.
In this technique, the user first obtains a path, i.e., diff, between old and new versions using an automated differencing technique.
Next, the user reads the diff and provides feedback indicating the unfavorable points, i.e., points that do not match with the user's intention.
In this figure, lines $A$ and $B$ were associated correctly; however, line $C$ was not associated and considered the deletion and insertion.
When the user thinks that the isolation of $C$ is inappropriate and points out it, the edges in the edit graph are updated, and the user obtains an updated diff based on the updated graph.
The orange-colored thick edges in the graph represent the shortest path in the first analysis, whereas the blue-colored thick edges represent the shortest path after the correction.
By updating the edit graph, the new shortest path follows, avoiding the use of the removed edges (represented as dashed edges) by the update.
Thus, it includes the mapping of the line $C$ rather than the lines $A$ and $B$.
The updated shortest path is automatically calculated based on the feedback using the same differencing algorithm.

The proposed approach is designed to provide users feedback on areas of dissatisfaction with the generated diff, rather than giving them feedback about the information on how the ideal diff will be.
This approach allows users who do not necessarily have the ideal diff in their mind to reach the ideal one by simply expressing their dissatisfaction with the current result.

\subsection{Updating Edit Graph}\label{s:feedback}

We provide three types of feedback actions: \emph{mismatch}, \emph{old-orphan}, and \emph{new-orphan}.
Feedback actions can be expressed as a pair of two indices with a special wildcard symbol: $a \in A\,(\subset (\mathbb{N}^+ \cup \{*\})^2)$.

\Heading{Mismatched Lines}
Consider a scenario where the $j$-th line in the new version is associated with $i$-th line in the old version (See \cref{f:mismatch}).
If the user believes that this scenario is inappropriate, the user indicates that they were undesired \emph{mismatched} lines, providing a feedback action of $\Fix{i}{j} \in A$.
The user desires either of the following cases for the target line in the new version:
  1) another mapping to another line in the old version, or
  2) no mapping to the lines in the old version, and it should be regarded as being inserted.
Therefore, we update the graph by removing the indicated diagonal edge as
\begin{align*}
   E \gets E \setminus \{ \Edge{i-1}{j-1}{i}{j} \}.
\end{align*}

Due to this removal of the diagonal edge in \cref{f:mismatch}, the original orange-colored path becomes infeasible, and the new blue-colored path is newly obtained by running the automated differencing algorithm again.

\Heading{Old-Orphan Lines}
Consider a scenario where the $i$-th line in the old version is not associated with any line in the new version (See \cref{f:old-orphan}).
If the user believes that this scenario is inappropriate, the user indicates that this was an undesired \emph{old-orphan} line, providing a feedback action of $\Fix{i}{*} \in A$.
The user should consider the disagreement of not only the current mapping but also any mappings that the old target line is not associated with any of the lines in the new version.
Therefore, we update the graph by removing all the horizontal edges that $i$-th line is regarded as deleted, as follows:
\begin{align*}
   E \gets E \setminus \{ \Edge{i-1}{j}{i}{j} \mid 0 \leq j \leq M \}.
\end{align*}

\Heading{New-Orphan Lines}
Consider a scenario where the $j$-th line in the new version is not associated to any line in the old version~(See \cref{f:new-orphan}).
If the user believes that this scenario is inappropriate, the user indicates that this was an undesired \emph{new-orphan} line, providing a feedback action of $\Fix{*}{j} \in A$.
The user should consider the disagreement of not only the current mapping but also the mappings that the target new line is not associated with any of the lines in the old version.
Therefore, we update the graph by removing all the vertical edges that $j$-th line is regarded as inserted, as follows:
\begin{align*}
 E \gets E \setminus \{ \Edge{i}{j-1}{i}{j} \mid 0 \leq i \leq N \}.
\end{align*}

\begin{figure*}[tb]\centering
  \includegraphics[width=0.8\linewidth]{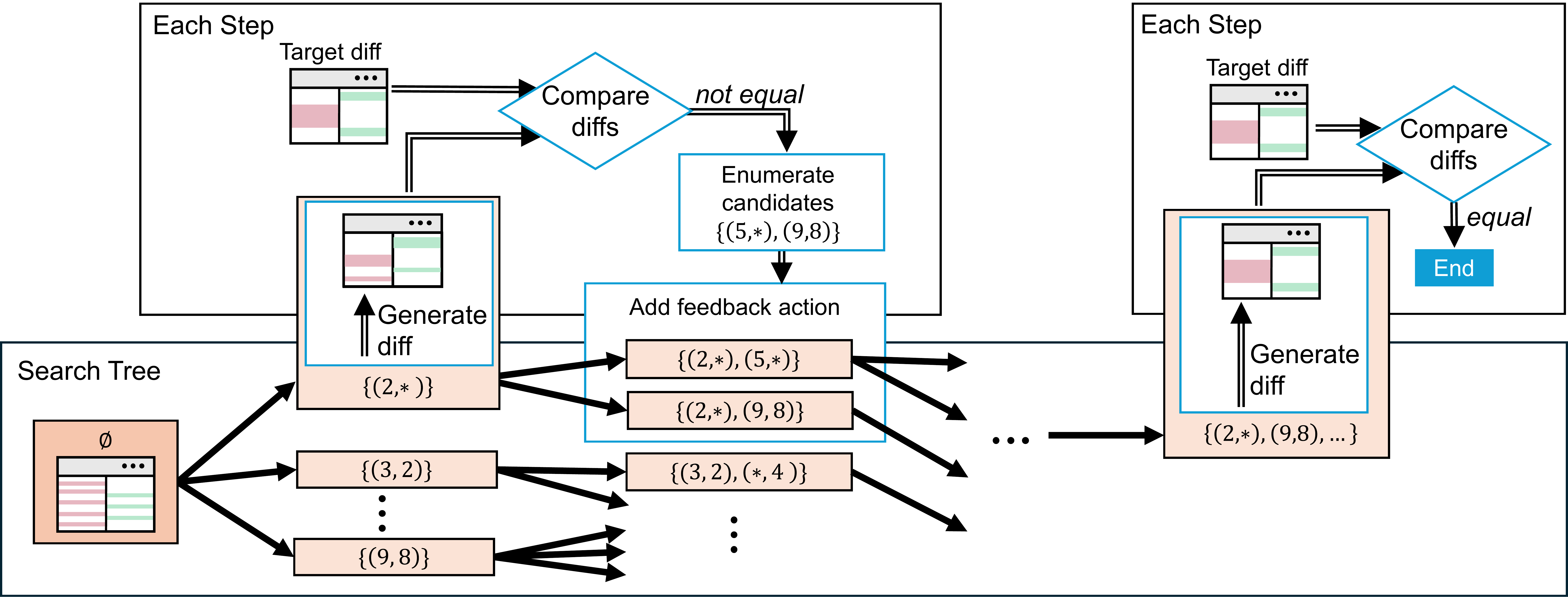}
  \caption{Overview of the automated simulation.}\label{f:overview}
\end{figure*}

The interactive optimization mechanism can be implemented through a simple user interface that allows users to click on a line of the displayed diff.
Any line on a diff, which can be added, deleted, or matched in the two versions, corresponds to an edge on the edit graph to be used to generate the diff.
Therefore, clicking on a line on the diff specifies an edge in the graph, with the intention of the dissatisfaction with the treatment of the specified line.
Then, a feedback action can be generated from the specified edge to update the diff.
The following $\Action \colon E \to A$ provides a feedback action from the specified edge:
\begin{align*}
   \Action(\Edge{i'}{j'}{i}{j}) = \begin{cases}
     (i,j) & \mathrm{if~} i' \neq i \land j' \neq j \text{~(diagonal)}, \\
     (i,*) & \mathrm{if~} i' = i \text{~(vertical)}, \\
     (*,j) & \mathrm{if~} j' = j \text{~(horizontal)}.
   \end{cases}
\end{align*}

After applying the above corrections, we calculate the paths again using the updated edit graph and obtain the new diff based on the calculated paths.

\section{Simulation Methodology}\label{s:sim}
\subsection{Objective}

Interactive diff optimization is helpful because it allows users to update diff as intended.
However, unlike existing automatic methods, this method requires user feedback.
Therefore, not only the quality of diffs generated by the method but also the effort required to optimize diff needs to be considered to discuss the practicality of this approach.

Although it is necessary to conduct a developer study for directly quantifying the effort, we have not yet gained an understanding of the fundamental properties of the interactive optimization approach.
Therefore, as a preliminary step, this paper investigates the ability of the proposed method to optimize the diff using two metrics: the number of feedback actions required to obtain optimal diff, and the number of simultaneously corrected spots by a feedback instance.
In interactive diff optimization, one feedback instance can correct multiple nonoptimal locations simultaneously.
Further, if one feedback action optimizes more inappropriate locations, it can be assumed that the overall effort to optimize diff will be more minor.

We can obtain the two metrics by mechanically simulating the interactive optimization process.
We concretely investigate these easily quantifiable optimization performances instead of employing approaches are costly to quantify.
For the investigation, we answer two research questions concerning different feedback strategies.

\def\RQone{What is the optimization performance in the ideal case?}
\def\RQtwo{What is the optimization performance in average cases?}

\textbf{\RQ{1}: \RQone}
In the first strategy, users provide ideal feedback and minimize the number of iterations.
In this case, the objective is measuring the highest performance of the interactive diff optimization method.
This study provides information on the extent to which the required effort for feedback can be reduced with this method, thereby making it a meaningful stepping stone for discussions.

\textbf{\RQ{2}: \RQtwo}
In scenarios where users utilize the tool, they may provide ideal feedback only occasionally.
Therefore, it is necessary to consider various feedback cases for evaluating the effort required for this method in a practical manner.
The average cases of feedback are investigated to examine these variations.

\subsection{Basic Design}

In the original interactive method, users optimize a diff by checking it and repeatedly providing feedback on the nonoptimal parts.
This repetitive process is formulated as a search problem for empirical study, which allows it to be simulated in a mechanical manner.

\Cref{f:overview} shows an overview of the simulation methodology.
This figure represents the process from the initial diff shown in \cref{f:badDiff}, iterating through feedback until obtaining the diff shown in \cref{f:goodDiff}.
The simulation is divided into two major parts: one part simulates the flow from checking the diff to providing feedback, as shown on the upper side of the figure, while the other part simulates various patterns of providing feedback, as shown on the lower side of the figure.

The following subsections of this section discuss the elements necessary for formulating a search problem using the optimization for the diffs shown in \cref{f:diff} as an example.

\subsection{Diff Generation Reflecting Feedback}\label{s:fixDiff}

The process of generating a diff was simplified in the simulation.
Instead of being incrementally updated via interactive feedback in the edit graph, it was updated through multiple simultaneous feedback per diff.
Users can recreate the same diff by reproducing the simultaneous feedback in sequence.
Thus, this simplification does not affect the outcome of the investigation.

\begin{figure}[tb]\centering
  \includegraphics[width=\columnwidth]{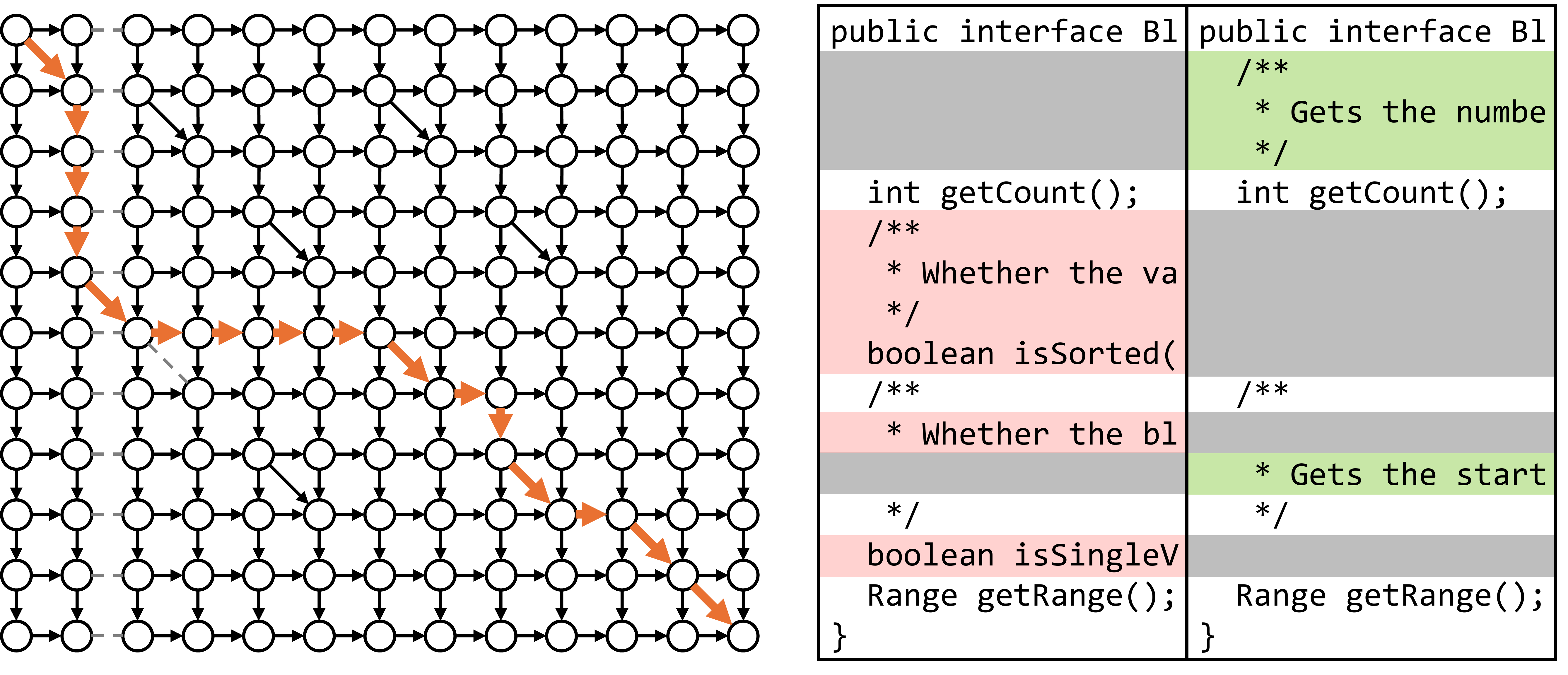}
  \caption{Diff generation reflecting the feedback.}\label{f:fixDiff}
\end{figure}

\begin{figure}[t]\centering
  \includegraphics[width=\linewidth]{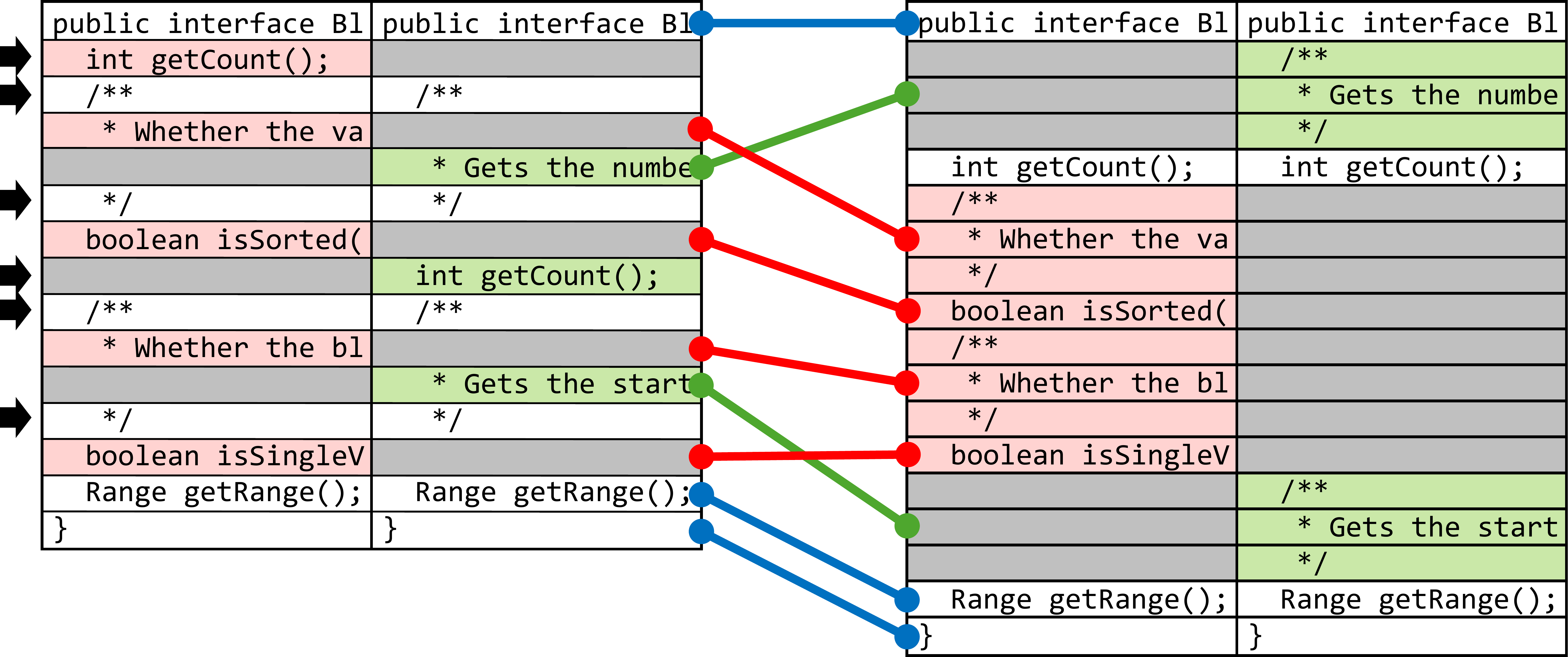}
  \caption{Generation of feedback action candidates.}\label{f:candidate}
\end{figure}

\Cref{f:fixDiff} demonstrates an example where the simulator provides feedback simultaneously for the removal of \texttt{getCount} on Line 2 of the old version and the matching of \texttt{/**} on Lines 3 and 6 of the old and new versions, respectively.
The corresponding edges in the edit graph are removed based on this feedback.
Consequently, the new diff produced differs from the diff shown in \cref{f:badDiff} before the feedback is given.

We formulated the process of generating a diff reflecting feedback as a feedback-aware differencer function $\Dfix \colon 2^A \to D$.
This function receives zero or more feedback actions and returns a path (a diff).
The feedback actions passed to the function in \cref{f:fixDiff} are $\{\Fix{2}{*},\Fix{3}{6}\}$.
The dashed edges on the edit graph are removed as per the provided feedback.
The differencer searches for the shortest path on this editing graph, and the path represented by the thick arrow indicates the value of $\Dfix(\{\Fix{2}{*},\Fix{3}{6}\})$.

\subsection{Target Diff}\label{s:target}

Defining the goal of the search is necessary for formulating the search problem.
Therefore, in the simulation, a different diff from the original one is predetermined, and the goal is set to generate a diff identical to this through repeated feedback.
This is referred to as the target diff, denoted by $d^* \in D$.

\Cref{f:goodDiff} represents the target diff.
Defining the target diff enables us to mechanically imitate the human process of repeatedly obtaining a diff from the diff shown in \cref{f:badDiff}, which are without feedback, until reaching the diff shown in \cref{f:goodDiff}.

\subsection{Generation of Feedback Candidates}\label{c:candidate}

For the simulation, we describe a method that mechanically calculates feedback actions for the generated current diff.
Parts needing feedback are where edit scripts differed when comparing the current and target diff.
The simulator creates feedback that can correct these points.
The simulator calculates a set of possible feedback candidates to provide because such feedback often exists in multiples.

\Cref{f:candidate} shows the comparison results between the diff in \cref{f:badDiff} and the target diff in \cref{f:goodDiff}.
Same edit scripts, such as the match between Line 1 of the old version and Line 1 of the new version with \texttt{public interface Bl$\dots$}, are connected by lines in both diffs.
Six edit scripts indicated by arrows in the current diff do not share any edit scripts with the target diff, and therefore, feedback actions applied to these six points become the next candidates to be performed.

Let $d \in D$ and $d^* \in D$ represent the diff being compared and the target diff, respectively.
We defined the feedback candidate function $c \colon D \times D \to 2^A$ that returns a set of feedback to be provided by comparing the two diffs as
\begin{align*}
  c(d; d^*) = \{ \Action(e) \mid e \in d \} \setminus
              \{ \Action(e) \mid e \in d^* \}.
\end{align*}

The number of feedback candidates $|c(d; d^*)|$ represents \textit{distance} between the current diff $d$ and target diff $d^*$.
Therefore, a diff with a shorter distance to the target diff is similar to the target diff.
In contrast, a diff with a longer distance to the target diff differs in many parts from the target one.

The number of feedback candidates $|c(d; d^*)|$ indicates how close the diff $d$ is to the target diff $d^*$.
In other words, the two diffs are closer if this value is small, and they are more distinct if this value is large.

In the following, we will refer to this value as \textit{similarity distance}.\footnote{The similarity distance is not a mathematical distance.
For example, symmetry does not always hold.}
Basically, the second argument is fixed as the target diff in that problem, and it is used as an index to represent the goodness of each diff when compared with the target diff.

\subsection{Search Problem}
\subsubsection{Formulation of Search Problem}

For the simulation, we described the interactive diff optimization as a search problem.
In each problem, an edit graph $(V, E)$, a universal set of possible feedback actions on the edit graph $A$, and a target diff $d^*$ are given.
Based on the conditions provided and preparations described in this section, we define the search state $s$, initial state $s_0$, goal condition $g$, and transition function $\delta$.

\Heading{Search State}
Each state in the search problem is expressed as a set of feedback actions to be given: $s \in S\,(\subseteq 2^A)$.
In \cref{f:overview}, each rectangle within the search tree represents a search state.
The search state represents a model of diff based on feedback in interactive diff optimization.

We did not use a diff as a search state because a different set of feedback actions can generate the same diffs.
Moreover, when the previous feedback actions are different, the resulting diffs may vary even if the same feedback is given for the same diff.
Therefore, we use a set of feedback actions as a search state.

\Heading{Initial State}
The initial state $s_0$ is defined as the state where the set of feedback actions is empty, specifically, $s_0 = \emptyset$.
The initial state models the first diff output in actual use.
If the set is empty, there are no changes in the edit graph, and the generated diff is the same.
Therefore, it is rational to consider the empty set as the initial state.
In \cref{f:overview}, the initial state is positioned at the left as the root of the search tree.

\Heading{Goal Condition}
The goal condition $g \colon S \to \{\top,\bot\}$ is a predicate that takes a search state and returns true if and only if the feedback-reflected diff matches the target diff.
The goal condition is formulated using the feedback-aware differencer function $\Dfix$, search state $s$, and target diff $d^*$ as 
\begin{align*}
  g(s) = \Dfix(s) \equiv d^*.
\end{align*}

If the state $s$ satisfies $g(s) = \top$, the simulator outputs that search state and terminates the search.
This corresponds to a use case where a user can obtain an optimal diff through interactive feedback $s$.

\Heading{Transition Function}
\def\Act{\mathit{succ}}
The actions applicable for each state add a single feedback action.
Therefore, action $a$ can be expressed as feedback action, where $a \in A$.
We defined a function $\Act \colon S \to 2^A$ that returns the set of applicable actions to bring state $s$ closer to target diff $d^*$.
This can be described using the feedback-aware differencer function $\Dfix$ and feedback candidate function $c$ as 
\begin{align*}
\Act(s) = c(\Dfix(s); d^*).
\end{align*}

Then, the transition function $\delta \colon S \times A \to S$, given state $s$ and action $a$, generates the next state as 
\begin{align*}
\delta(s,a) = s \cup \{a\}.
\end{align*}
This represents one iteration of the interactive method where users provide feedback on the nonoptimal points of the diff.

\subsubsection{Formulation of the Search Space}
The search space $S$ is defined based on the aforementioned formulation.
First, the initial state is included in the search space, i.e., $s_0 \in S$.
The upper side of \cref{f:overview} represents the flow where the child states of a state $s \in S$ are added to the search space.
First, a diff $\Dfix(s)$ is generated based on feedback, and then, the goal condition $g(s)$ is checked.
If true, the search ends, and there are no child states; if false, the creation of child states continues.
Based on the diff, the possible set of actions $\Act(s)$ is calculated.
Finally, new states can be created from the state and the action.
All these can be the states of the search space:
\begin{align*}
s \in S \Rightarrow g(s) \,\lor\, \forall a(a\in \Act(s) \Rightarrow \delta(s,a) \in S).
\end{align*}
The search space was defined based on the initial state and this addition process.

The example in \cref{f:overview} demonstrates the process of adding child states of state $\{\Fix{2}{*}\}$.
The differencer generates diff $\Dfix(\{\Fix{2}{*}\})$; however, it differs from the target diff $d^*$.
Consequently, the child states $\{\Fix{2}{*},\Fix{5}{*}\}$ and $\{\Fix{2}{*},\Fix{9}{8}\}$ are added to the search tree.

\section{Empirical Study}\label{s:empirical}
We simulated interactive diff optimization in the format of the search problems written in the previous section.
Based on the simulation, we empirically investigated the optimization performance of the proposed method.
For the empirical study, we considered two types of feedback strategies: (1) providing ideal feedback and (2) providing random feedback.
The corresponding \RQ{}s were established to evaluate the optimization performance of these two feedback strategies.

\subsection{Data Collection}
The manual preparation of the target diffs was challenging because this investigation requires many source code changes and the corresponding target diffs.
Therefore, in the empirical investigation, we created the target diffs using a Histogram algorithm.
In interactive diff optimization, diffs are generated based on the shortest path search of edit graphs.
The concept prioritized in the Histogram algorithm differs from this approach, and therefore, the diffs are dissimilar within a practical range.
As such, this enables the efficient preparation of the required target diffs that differed from the ones without the feedback.

Previous research \cite{Nugroho2020} investigated diffs generated by the Histogram algorithm.
The source code changes where the diff generated by Histogram and Myers algorithms differ were published as artifacts from 24 GitHub projects.
In this study, we preliminarily investigated whether diffs generated using the Histogram algorithm differed from those generated by our differencer on this dataset.
We used the changes that created different diffs as the dataset.

Furthermore, we filtered for problems where lines of code were 3,000 or less in both the new and old versions and the number of feedback candidates in the initial state was 30 or less to limit the size of the search problem and computation time.
In addition, the same changes about a license update existed in more than 700 files, and therefore, these files were excluded from the dataset.
Thus, we used 9,229 source code changes from 23 projects that met the above criteria as our dataset.
Table \ref{t:dataset} summarizes these 9,229 changes for the lines of code in the new and old versions, number of changed lines, and number of initial feedback candidates.
\begin{table}[tb]\centering
  \caption{Attribute of Dataset}\label{t:dataset}
  {\tabcolsep=4.5pt\begin{tabular}{lrrrrrr} \hline
      & Min & Q1 & Q2  & Q3 & Max & Average\\\hline
    LOC in the old version & 8 & 138 & 285 & 618 & 2,929 & 493.73\\
    LOC in the new version & 3 & 153 & 310 & 654 & 2,993 & 515.34\\
    \# changed lines & 3 & 55 & 111 & 223 & 3,779 & 182.29 \\
    initial similarity distance & 2 & 2 & 5 &11 & 30 & 7.78\\\hline
  \end{tabular}}
\end{table}

Additionally, when conducting the simulations, blank lines, consisting solely of the newline character, were removed as a preprocessing step.
It is unlikely for users to optimize for blank lines because information about how the blank line is changed is rarely useful to the user in understanding the change.
However, states do not satisfy the goal condition until they fully match the target diff in the simulation, and therefore, there was considerable feedback on optimizing blank lines.
If such cases were included in the study, the results would be impractical.
Therefore, eliminating unnecessary feedback related to blank lines was necessary to obtain more practical research outcomes.

\subsection{\RQ{1}: \RQone}
\subsubsection{Motivation}
The optimization performance of the proposed method is the highest in the case where the ideal feedback is provided, i.e., when the number of feedback actions is minimized.
This case demonstrates the most significant benefit of interactive diff optimization, and investigating these results is meaningful in considering the effort involved in this interactive method.
Therefore, in \RQ{1}, we investigated the optimization performance of the feedback in the ideal case and the factors that influence it.

\subsubsection{Study Design}
The simulator applied the A* algorithm to a search problem that imitated diff optimization and found a state that satisfied the goal condition with the fewest number of feedback actions.
The subject of this study was 9,229 source code changes in the dataset, with a maximum execution time limit of 30 min.

\begin{figure}[tb]\centering
  \includegraphics[scale=0.11]{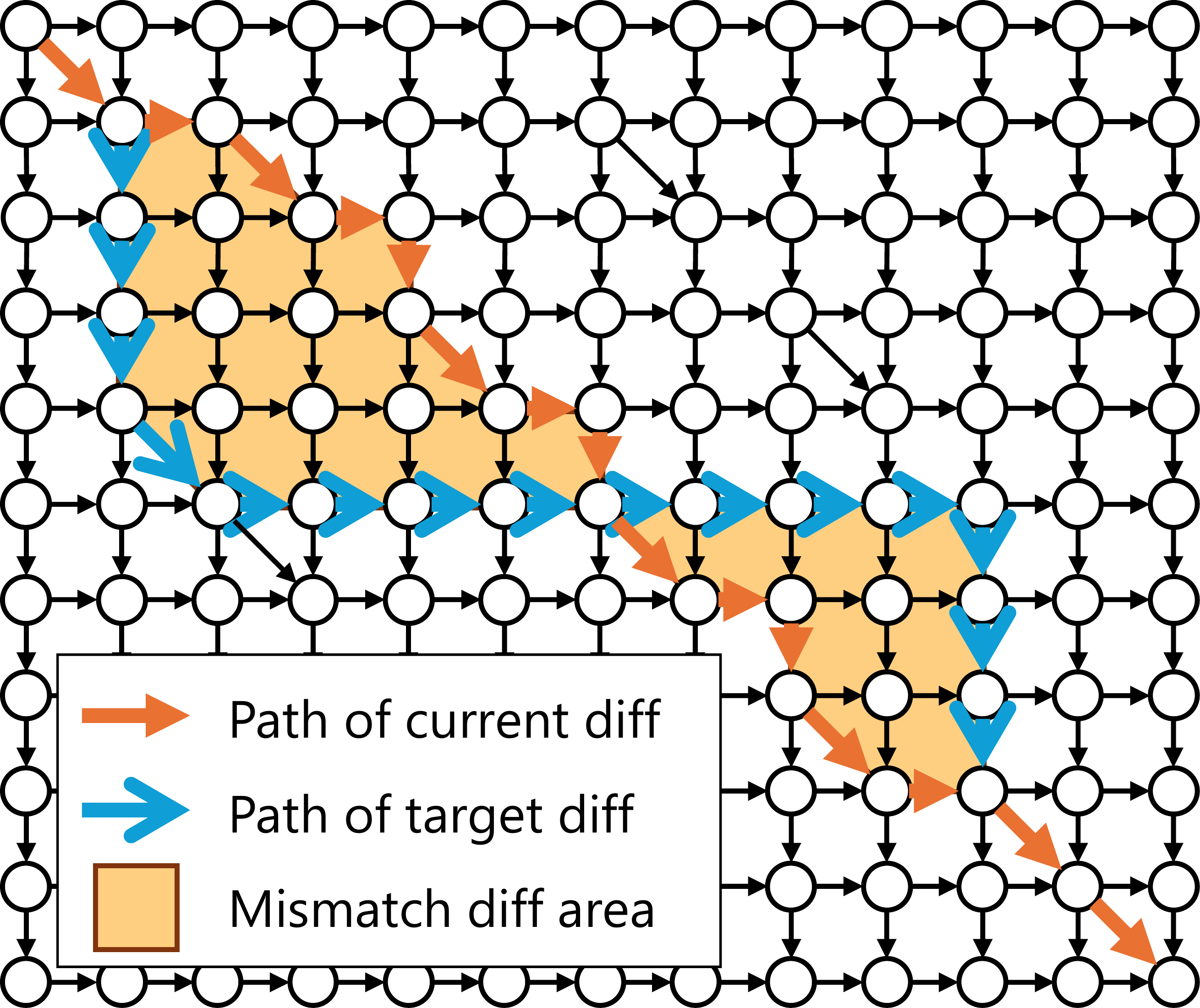}
  \caption{Mismatch diff area.}\label{f:misamatchArea}
\end{figure}

The A* search requires a heuristic function, for which we defined mismatch diff areas.
These are continuous areas on the edit graph, as illustrated by the two yellow areas in \cref{f:misamatchArea}, enclosed by the current and target diffs.
The number of feedback instances required to optimize diff must equal or exceed the number of mismatch diff areas.
In the example shown in the figure, users must provide feedback at least twice.
This fact can be stated for the following reasons: 
(1) Edges on different mismatch diff areas are not removed in a single feedback action because edges removed in feedback are either single or multiple edges parallel to the vertical or horizontal, whereas the mismatched diff areas are arranged diagonally.
(2) If part of the current diff and target diff share the same start and end nodes, the partial path between the start and end nodes in the current diff always takes precedence because the differencer stably prioritizes paths when the starting and ending nodes of two paths are the same to prevent variations in the results of each simulation, and the prioritized path is the current diff.

Thus, each area must have its edges removed through different feedback actions to obtain the target diff.
The number of areas acts as a heuristic function $h \colon S \to \mathbb{N}$.
\begin{align*}
  h(s) = |\{ \Pt{i}{j} \mid \, 
    & 0 \leq i,i',i'' \leq N \land 0 \leq j,j',j''\leq M \land {}\\
    & \Edge{i'}{j'}{i}{j} \in \Dfix(s) \land \Edge{i''}{j''}{i}{j} \in d^* \land {}\\
    & (i' \neq i'' \lor j' \neq j'') \}|
\end{align*}

To reduce the overall time taken, we first use the similarity distance $|c(\Dfix(s), d^*)|$ as a simpler, efficient, but non-admissible heuristic function.
If the obtained search path is longer than the trivial optimal length, i.e., the number of mismatch diff areas for the initial diff, $h(s)$ is used as an admissible heuristic function to compute an optimal path.

\subsubsection{Results and Discussion}
\begin{figure}[tb]\centering
  \includegraphics[width=0.65\linewidth]{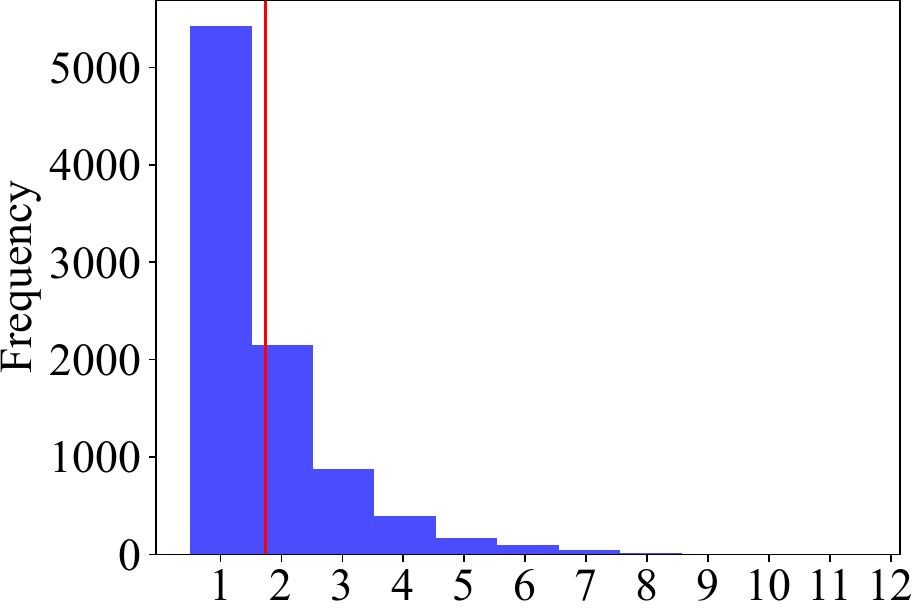}
  \caption{Distribution of the minimum number of feedback action.}\label{f:times}
\end{figure}

\begin{figure}[tb]\centering
    \includegraphics[width=0.65\linewidth]{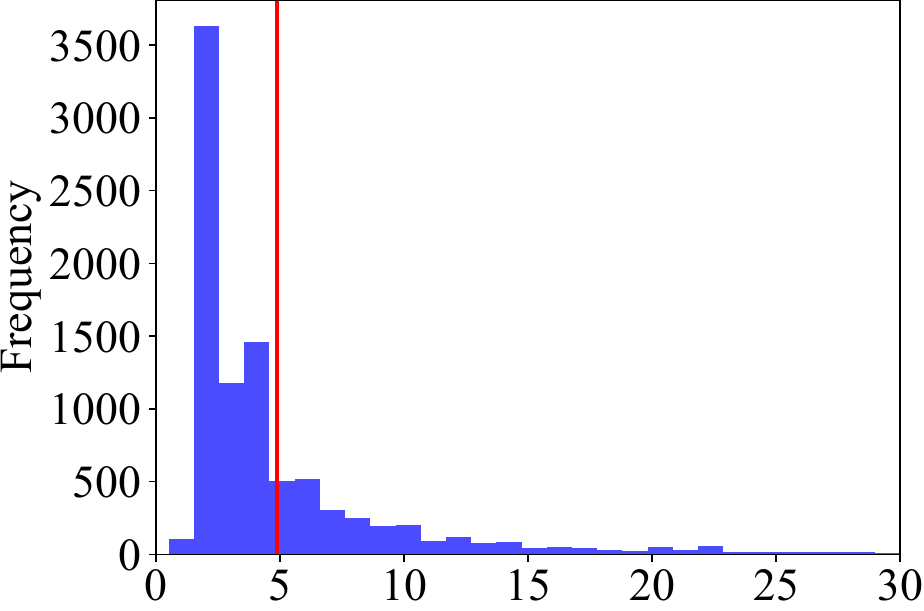}
    \caption{Distribution of the average speed.}\label{f:value}
\end{figure}

Excluding 44 changes that failed because of exceeded time limits or insufficient memory, 9,185 entries were used.
\Cref{f:times} shows the minimum number of feedback instances required to achieve the target diff.
\Cref{f:value} shows the \emph{average speed}, which represents the improvement of a single feedback action, and it is computed as the initial similarity distance divided by the minimum number of feedback actions.
The red lines in each graph represent the average values of the data.

In \cref{f:times}, the average of the smallest number of feedback actions is 1.73.
In 59\% of the dataset, only one feedback action made the diffs equal to the target diff.
In 92\% of the cases, the diffs were corrected to the target diffs with three or fewer feedback actions.
Thus, the effort required to gain the target diff is considerably tiny in the ideal cases.

The average number of average speed was approximately 4.87 locations.
The mode was 2 in \cref{f:value} because the most common and simple pattern among initial diffs was the case where there were two feedback candidates.
Most of them were addressed in a single feedback action, and therefore, the value of 2 is prominently numerous in the graph.
Multiple parts can be addressed simultaneously in an ideal case because values greater than one are widely distributed in the graph, so the interactive method can efficiently optimize diffs.

\begin{figure}[tb]\centering
  \includegraphics[width=0.7\linewidth]{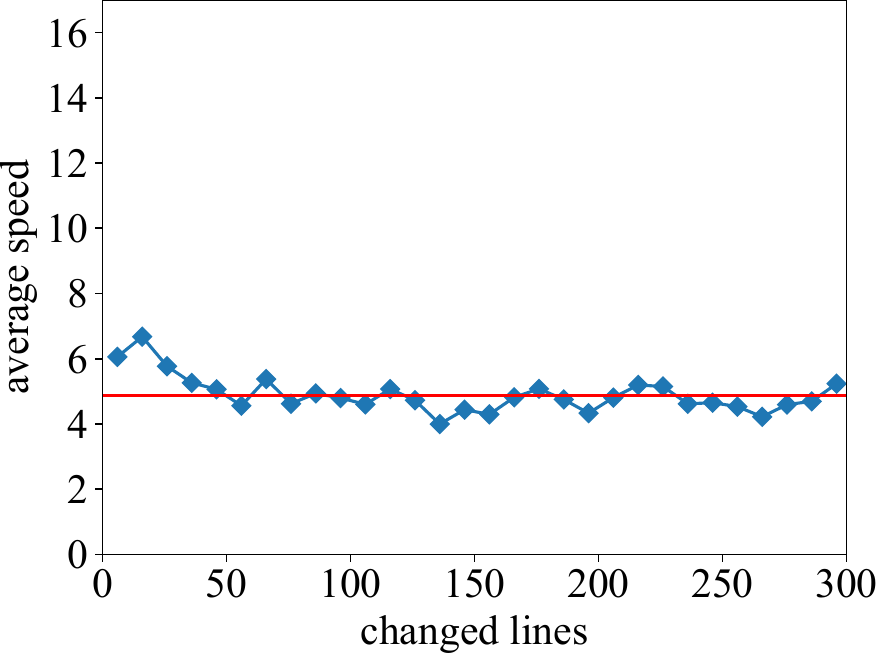}
  \caption{Average speed against the number of changed lines.}\label{f:change}
\end{figure}
\begin{figure}[tb]\centering
    \includegraphics[width=0.7\linewidth]{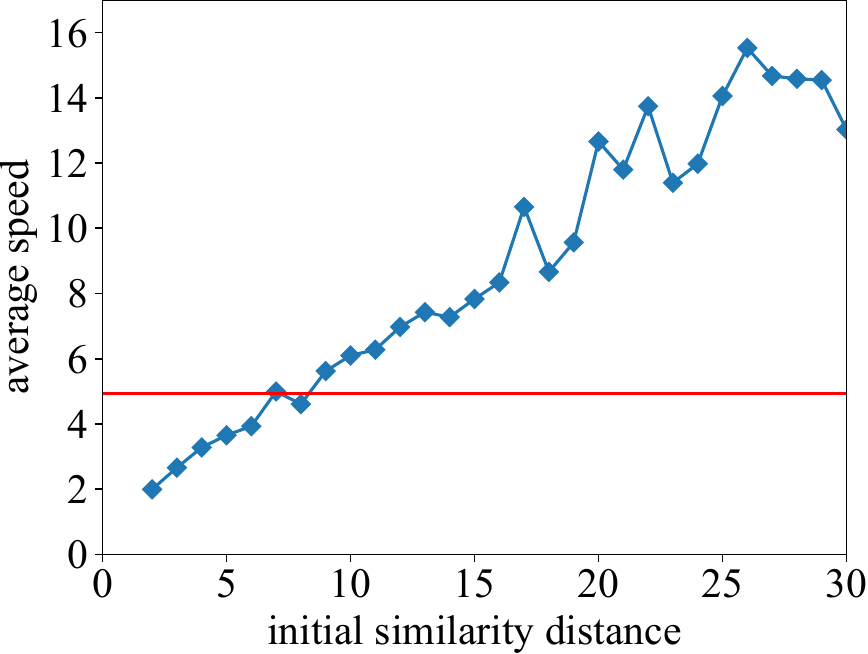}
    \caption{Average speed against initial similarity distance.}
    \label{f:sousa}
\end{figure}

As shown in \cref{f:change}, the average speed does not vary although the number of lines changed differs.
Therefore, according to feedback, the number of changed lines is not relevant to the optimization performance.

\Cref{f:sousa} shows the effect of the similarity distance of the initial diff on the optimization performance.
If the distance is large, more feedback is needed to achieve the target diff.
At the same time, when the initial distance was large, the average speed was also high.
Therefore, an increase in the number of feedback caused by the large number of nonoptimal parts could be suppressed, and such a diff required less effort.

\Conclusion{
The interactive optimization approach has shown its potential to optimize a diff with minimal effort in ideal case simulation.
In 92\% of the cases, three or fewer feedback actions were required to achieve the target diffs.
In addition, 4.87 lines were fixed per feedback action on average.
}

\subsection{\RQ{2}: \RQtwo}
\subsubsection{Motivation}
Users can provide ideal feedback only occasionally when optimizing diffs.
Considering not only the ideal cases but also the general ones is a rational approach to evaluate the performance of interactive diff optimization.
Since \RQ{1} investigated the case with the minimum number of indications, such typical cases were not included in the results.
\RQ{2} evaluates the diffs when various types of feedback are provided, and the results include multiple cases.

\subsubsection{Study Design}
We investigated how each feedback action in the initial state affected the diff.
The breadth first search was performed on the search problem, and the similarity distance of the diffs for all states at depth 1 was logged.

We calculated \emph{$\mathit{\Delta}$distance}, the reduction in similarity distance before and after feedback, which indicates how much the initial diff was brought closer to the target diff.
From there, the average, best, and worst values were recorded as representative values for each simulation.
Code changes for which results were not obtained in \RQ{1} were excluded from the results in answering this research question.

\subsubsection{Results and Discussion}

\begin{figure}[tb]\centering
  \includegraphics[width=\linewidth]{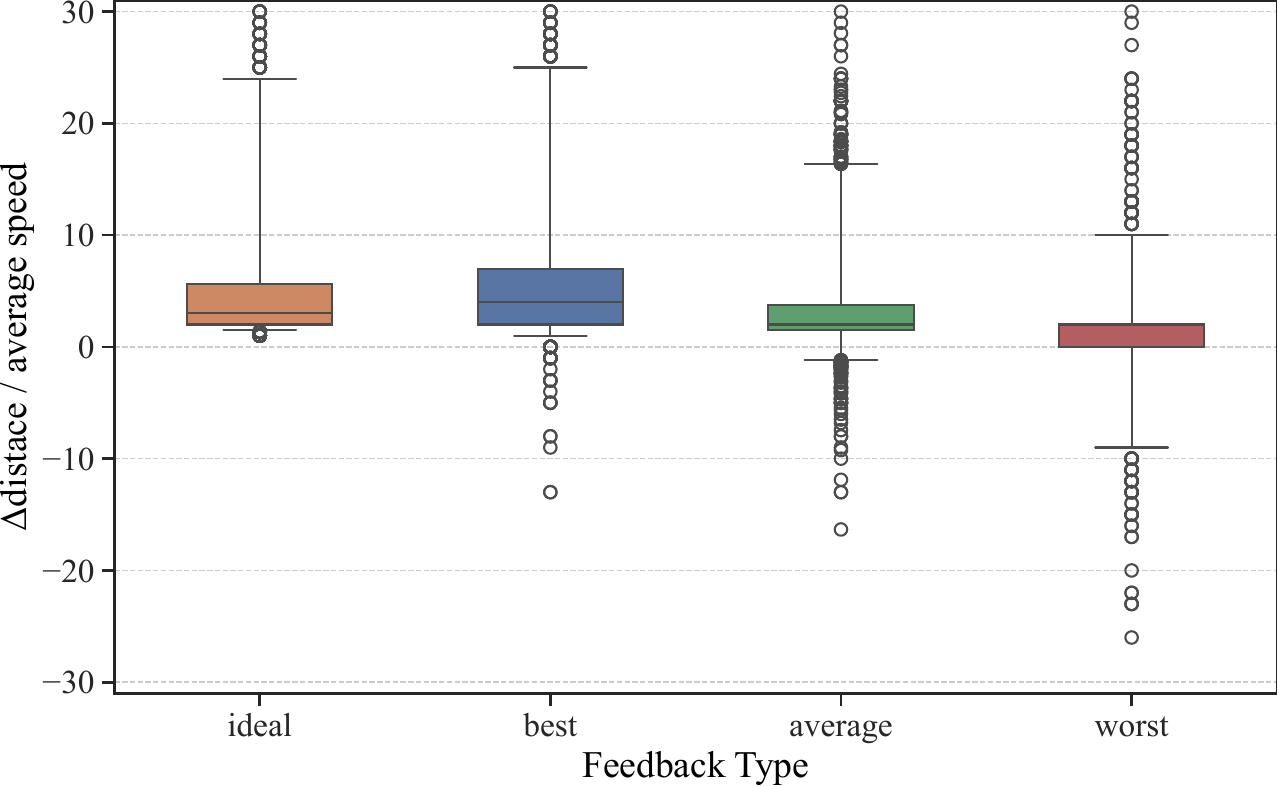}
  \caption{$\Delta$distance for each feedback pattern.}\label{f:rq2_box}
\end{figure}

\begin{table}[tb]
  \centering
  \caption{Average and Ratio to Average Speed for Each Type of Feedback in $\Delta$distance}
  \label{t:rq2_matrix}
  \begin{tabular}{lrrrr} \hline
      & Ideal & Best & Average  & Worst\\ \hline
    $\Delta$distance (average speed) & 4.87 & 5.67 & 3.14 & 0.71\\
    Ratio to average speed & 1 & 1.14 & 0.68 & 0.22\\ \hline
  \end{tabular}
\end{table}

\Cref{f:rq2_box} shows the $\Delta$distance for each category, and the average speed of the ideal case is added for reference.
\cref{t:rq2_matrix} shows the average per category and the ratio compared to the average speed.
The whiskers of the box plot are set to the lower 1\% and upper 99\%.
The ideal is represented as a box plot in \cref{f:value}.

In the best-case scenario, the distribution is longer above than in the ideal case, and the ratio exceeds 1.
This is because there are feedbacks that significantly approach the target difference and those that do not when the number of feedback instances is more than one.

In the best-case scenario, the $\Delta$distance is 5.67, whereas in the average case, it dropped to about 3.14, which is approximately 55\%.
The ratio between the average and ideal cases averaged 68\%.
Therefore, not all feedback is equally performed, and on average, the optimization performance is inferior to that in ideal scenarios.

The parts of the figure below zero represent cases where the diff obtained by feedback was farther from the target diff than the previous one.
15\% of the worst-case patterns fall into this category, and it was 2\% for the average patterns.
Therefore, while the diffs can worsen due to feedback, such situations rarely occur on average.

\Conclusion{
  The feedback in the average cases had a lower optimization performance than that in the ideal case.
  However, since more than three lines were fixed simultaneously to approach the target diff, the feedback effort in the average case is expected to be enough low.
}

\subsection{Threats to Validity}
For filtering the dataset, \RQ{1} showed different results from those of an unfiltered dataset because feedback candidates with a diff greater than 30 were filtered, and including them would increase both the number of feedback instances and the average speed.
However, attempting to include these would result in more simulation failures within the dataset, leading to unreliable research results, which is why the current experimental setup was adopted.

In \RQ{2}, we focused not on all the feedback within the search problem, but on those from the initial state.
In interactive diff optimization, the user checks the current diff and provides feedback; there is no need to worry about the past.
Therefore, examining the feedback that can be given in the initial state is sufficient to evaluate the performance.

Target diffs using the Histogram algorithm were generated mechanically in the empirical study.
Thus, they included diffs that were nonoptimal.
This investigation aimed to demonstrate the extent to which feedback affects the diffs quantitatively, and the quality of the diffs is not a concern.

\section{Related Work}\label{s:relatedwork}

A lot of differencing algorithms and tools have been proposed.
Among these, the most basic is a line-based approach.
Some of these algorithms use structural information such as XML~\cite{xdiff}.
For example, JDiff~\cite{jdiff} uses the structural information of the program source code, whereas Change Distiller~\cite{change-distiller} uses abstract syntax trees.
GumTree~\cite{gumtree} is another efficient differencing algorithm for source code based on the matching on syntax trees.

Semantic Diff~\cite{semantic-diff} applies LCS-based differencing to the sequences of program statements.
This method is useful both for efficiency and accuracy.
However, it produces incorrect mappings so that the approach of correcting differences remains effective.

Considering the correction or adaptation of differences of software artifacts is not novel, and it has already been discussed previously \cite{sidiff,user-driven-adaption}.
The main difference of the proposed approach is that it focuses on interactive correction using a semi-automated tool that enables developers to update the diffs without specifying the ideal mapping.

Constrained longest common subsequence~(C-LCS) is a sub-category of LCS having constraints.
One example of the constraints is that the resulting LCS should include the given particular subsequence \cite{clcs-variant,clcs}.
The corrections of an edit graph in the proposed technique can be regarded as constraints.
However, these work do not mention about the application to the computation of source code diffs.

\section{Conclusion}\label{s:conclusion}

In this study, we proposed an interactive optimization of source code differences, a method that creates optimal differences crucial for understanding changes during code reviews, which is an essential process in software development.
Investigating the effort to evaluate practicality is necessary because the proposed method requires human feedback.
To gain insights into the effort as a preliminary stage of the subject experiment, we empirically investigated the optimization performance of feedback.
In this investigation, the interactive process was formulated and simulated as a search problem.

As a result, in the ideal case, feedback was required on average 1.73 times.
Furthermore, an average of 4.87 edit scripts were corrected simultaneously with one feedback instance.
The effectiveness of feedback in average cases was about 68\% compared to ideal cases.

Future challenges can be listed up as follows:

\Heading{Simulation of More Diverse Data}
In this study, differential generation in the simulation employs dynamic programming.
Therefore, the generation of differences was time-consuming, and in \RQ{1}, it was necessary to impose filter conditions to restrict the search space.
Accelerating the simulation can be expected to yield results for more complex problems and allow for deeper examination of optimization performance.

The Histogram algorithm used for generating the dataset does not imitate all practical usage scenarios.
Therefore, creating simulation datasets based on different approaches would also be useful in achieving comprehensive results.
For example, it may be possible to constructing more precise oracles of optimal diffs based on the records obtained by monitoring the interaction between developers and development environments\cite{Negara2012,Maruyama2018,Yoon2011}.

\Heading{Manual Assessment}
Information on the feedback provided and the resulting difference improvements can be obtained through the simulation.
Humans can check and evaluate ideal or low-performance feedback.
Thus, it is conceivable to improve the method so that humans can optimize more through natural feedback.

\Heading{Optimization across Different Granularities}
The current method generates and optimizes the differences between source code elements line-by-line.
As a future extension, the proposed method could optimize differences at granularities other than lines, such as at the token or syntactic levels.
Such extensions will enable more appropriate differences that better reflect the intentions of the reviewers.

Supplemental material, including the experimental results, is available \cite{dataset}.

\section*{Acknowledgments}
This study was partly supported by JSPS Grants-in-Aid for Scientific Research Nos.\ JP24H00692, JP23K24823, JP21H04877, JP21K18302, and JP21KK0179.

\IEEEtriggeratref{13}
\bibliographystyle{IEEEtran}
\bibliography{references} 

% Generated by IEEEtran.bst, version: 1.12 (2007/01/11)
\begin{thebibliography}{10}
\providecommand{\url}[1]{#1}
\csname url@samestyle\endcsname
\providecommand{\newblock}{\relax}
\providecommand{\bibinfo}[2]{#2}
\providecommand{\BIBentrySTDinterwordspacing}{\spaceskip=0pt\relax}
\providecommand{\BIBentryALTinterwordstretchfactor}{4}
\providecommand{\BIBentryALTinterwordspacing}{\spaceskip=\fontdimen2\font plus
\BIBentryALTinterwordstretchfactor\fontdimen3\font minus \fontdimen4\font\relax}
\providecommand{\BIBforeignlanguage}[2]{{%
\expandafter\ifx\csname l@#1\endcsname\relax
\typeout{** WARNING: IEEEtran.bst: No hyphenation pattern has been}%
\typeout{** loaded for the language `#1'. Using the pattern for}%
\typeout{** the default language instead.}%
\else
\language=\csname l@#1\endcsname
\fi
#2}}
\providecommand{\BIBdecl}{\relax}
\BIBdecl

\bibitem{McIntosh2015}
S.~McIntosh, Y.~Kamei, B.~Adams, and A.~E. Hassan, ``An empirical study of the impact of modern code review practices on software quality,'' \emph{Empirical Software Engineering}, vol.~21, pp. 2146--2189, 2016.

\bibitem{Bacchelli2013}
A.~Bacchelli and C.~Bird, ``Expectations, outcomes, and challenges of modern code review,'' in \emph{Proceedings of the 35th International Conference on Software Engineering (ICSE 2013)}, 2013, pp. 712--721.

\bibitem{Yida2012}
Y.~Tao, Y.~Dang, T.~Xie, D.~Zhang, and S.~Kim, ``How do software engineers understand code changes? {A}n exploratory study in industry,'' in \emph{Proceedings of the 20th International Symposium on the Foundations of Software Engineering (FSE 2012)}, ser. FSE '12, 2012, pp. 1--11.

\bibitem{Nugroho2020}
Y.~S. Nugroho, H.~Hata, and K.~Matsumoto, ``How different are different diff algorithms in {Git}?: Use \verb|--histogram| for code changes,'' \emph{Empirical Software Engineering}, vol.~25, pp. 790--823, 2020.

\bibitem{Ram2018}
A.~Ram, A.~A. Sawant, M.~Castelluccio, and A.~Bacchelli, ``What makes a code change easier to review: {A}n empirical investigation on code change reviewability,'' in \emph{Proceedings of the 26th ACM Joint Meeting on European Software Engineering Conference and Symposium on the Foundations of Software Engineering (ESEC/FSE 2018)}, 2018, pp. 201--212.

\bibitem{Barnett2015}
M.~Barnett, C.~Bird, J.~Brunet, and S.~K. Lahiri, ``Helping developers help themselves: Automatic decomposition of code review changesets,'' in \emph{Proceedings of the 37th International Conference on Software Engineering (ICSE 2015)}, vol.~1, 2015, pp. 134--144.

\bibitem{Canfora2009}
G.~Canfora, L.~Cerulo, and M.~Di~Penta, ``{Ldiff}: {A}n enhanced line differencing tool,'' in \emph{Proceedings of the 31st International Conference on Software Engineering (ICSE 2009)}, 2009, pp. 595--598.

\bibitem{Sadowski2018}
C.~Sadowski, E.~S\"{o}derberg, L.~Church, M.~Sipko, and A.~Bacchelli, ``Modern code review: {A} case study at {Google},'' in \emph{Proceedings of the 40th International Conference on Software Engineering: Software Engineering in Practice (ICSE-SEIP 2018)}, 2018, pp. 181--190.

\bibitem{Myers1986}
E.~W. Myers, ``An {O(ND)} difference algorithm and its variations,'' \emph{Algorithmica}, vol.~1, no.~1, pp. 251--266, 1986.

\bibitem{patch-flow}
P.~C. Rigby, D.~M. German, and M.-A. Storey, ``Open source software peer review practices: A case study of the {Apache} server,'' in \emph{Proceedings of the 30th International Conference on Software Engineering (ICSE 2008)}, 2008, pp. 541--550.

\bibitem{lcs-diff}
W.~Miller and E.~W. Myers, ``A file comparison program,'' \emph{Software: Practice and Experience}, vol.~15, no.~11, pp. 1025--1040, 1985.

\bibitem{xdiff}
Y.~Wang, D.~J.DeWitt, and J.-Y. Cai, ``{X-Diff}: An effective change detection algorithm for {XML} documents,'' in \emph{Proceedings of the 19th International Conference on Data Engineering (ICDE 2003)}, 2003, pp. 519--530.

\bibitem{jdiff}
T.~Apiwattanapong, A.~Orso, and M.~J. Harrold, ``A differencing algorithm for object-oriented programs,'' in \emph{Proceedings of the 19th IEEE International Conference on Automated Software Engineering (ASE 2004)}, 2004, pp. 2--13.

\bibitem{change-distiller}
B.~Fluri, M.~Wuersch, M.~Pinzger, and H.~Gall, ``Change distilling: Tree differencing for fine-grained source code change extraction,'' \emph{IEEE Transaction on Software Engineering}, vol.~33, no.~11, pp. 725--743, 2007.

\bibitem{gumtree}
J.-R. Falleri, F.~Morandat, X.~Blanc, M.~Martinez, and M.~Monperrus, ``Fine-grained and accurate source code differencing,'' in \emph{Proceedings of the 29th ACM/IEEE International Conference on Automated Software Engineering (ASE 2014)}, 2014, pp. 313--324.

\bibitem{semantic-diff}
A.~Yoshida, S.~Yamamoto, and K.~Agusa, ``Semantic diff,'' \emph{IPSJ Journal}, vol.~38, no.~6, pp. 1163--1171, 1997.

\bibitem{sidiff}
T.~Kehrer, U.~Kelter, P.~Pietsch, and M.~Schmidt, ``Adaptability of model comparison tools,'' in \emph{Proceedings of the 27th IEEE/ACM International Conference on Automated Software Engineering (ASE 2012)}, 2012, pp. 306--309.

\bibitem{user-driven-adaption}
K.~M\:{u}ller and B.~Rumpe, ``User-driven adaptation of model differencing results,'' in \emph{Proceedings of the International Workshop on Comparison and Versioning of Software Models (CVSM 2014)}, 2014.

\bibitem{clcs-variant}
P.~Bonizzoni, G.~D. Vedova, R.~Dondi, and Y.~Pirola, ``Variants of constrained longest common subsequence,'' \emph{Information Processing Letters}, vol. 110, no.~20, pp. 877--881, 2010.

\bibitem{clcs}
Y.-T. Tsai, ``The constrained longest common subsequence problem,'' \emph{Information Processing Letters}, vol.~88, no.~4, pp. 173--176, 2003.

\bibitem{Negara2012}
S.~Negara, M.~Vakilian, N.~Chen, R.~E. Johnson, and D.~Dig, ``Is it dangerous to use version control histories to study source code evolution?'' in \emph{Proceedings of the 26th European Conference on Object-Oriented Programming (ECOOP 2012)}, 2012, pp. 79--103.

\bibitem{Maruyama2018}
K.~Maruyama, S.~Hayashi, and T.~Omori, ``{ChangeMacroRecorder}: Recording fine-grained textual changes of source code,'' in \emph{Proceedings of the 25th International Conference on Software Analysis, Evolution and Reengineering (SANER 2018)}, 2018, pp. 537--541.

\bibitem{Yoon2011}
Y.~Yoon and B.~A. Myers, ``Capturing and analyzing low-level events from the code editor,'' in \emph{Proceedings of the 3rd ACM SIGPLAN Workshop on Evaluation and Usability of Programming Languages and Tools (PLATEAU 2011)}, 2011, pp. 25--30.

\bibitem{dataset}
T.~Yagi and S.~Hayashi, ``Online appendix of ``{T}oward interactive optimization of source code differences: An empirical study of its performance'','' Zenodo, \url{https://doi.org/10.5281/zenodo.13618978}, 2024.

\end{thebibliography}

\end{document}